\documentstyle[aps,twocolumn,prb,epsf]{revtex}

\begin{document}
                                             
\newcommand{\be}{\begin{equation}}
\newcommand{\ee}{\end{equation}}
\newcommand{\beq}{\begin{eqnarray}}
\newcommand{\eeq}{\end{eqnarray}}
\newcommand{\lsim}{\:\raisebox{-0.5ex}{$\stackrel{\textstyle<}{\sim}$}\:}
\newcommand{\gsim}{\:\raisebox{-0.5ex}{$\stackrel{\textstyle>}{\sim}$}\:}

\draft

\twocolumn[\hsize\textwidth\columnwidth\hsize\csname @twocolumnfalse\endcsname

\title{Disordered Hubbard model in $d=\infty$}

\author{M. S. Laad${}^1$, L. Craco${}^2$, and E. M\"uller-Hartmann${}^1$}

\address{${}^1$Institut fuer Theoretische Physik, Universitaet zu Koeln, 
Z\"ulpicher Strasse, 50937 Koeln, Germany \\
${}^2$Instituto de F\'{\i}sica ``Gleb Wataghin'' - UNICAMP, 
C.P. 6165, 13083-970 Campinas - SP, Brazil}

\date{\today}

\maketitle

\widetext

\begin{abstract}  
We investigate the effects of static, diagonal disorder in the $d=\infty$ 
Hubbard model by treating the dynamical effects of local Hubbard 
correlations and disorder on an equal footing.  This is achieved by a proper
combination of the iterated perturbation theory and the coherent potential 
approximation.  Within the paramagnetic phase, we find that the renormalized 
Fermi liquid metal phase of the pure Hubbard model is stable against disorder 
for small disorder strengths.  With increasing disorder, strong resonant 
scattering effects destroy low-energy Fermi liquid coherence, leading to an
{\it incoherent} metallic non-FL state off half-filling.  Finally, for large 
enough disorder, a {\it continuous} transition to the Mott-Anderson insulating 
phase occurs.  The nature of the non-FL metallic phase, as well as the effects
of the low energy coherence (incoherence) on optical conductivity and 
electronic Raman spectra, are considered in detail. 

\end{abstract}
     

]

\narrowtext

\section{INTRODUCTION}
    
Interest in doped transition metal oxides, a longstanding problem of 
interest, has been renewed by discovery of non-Fermi liquid behavior observed
in the normal state of the high-$T_{c}$ cuprates, as well as certain other 
transition metal and rare-earth compounds.~\cite{a1}  In most cases, the 
non-Fermi liquid anomalies are observed on doping the Mott-Hubbard insulator, 
a process that is achieved by carrier doping.  In real systems, carrier 
doping is brought about by chemical substitution, a process that simultaneously
introduces {\it disorder} on some length scale.  Given this, the electrons 
``feel'' different local environments, and it is physically meaningful to speak
only of {\it disorder averaged} quantities.  Calculations carried out on models
like the pure Hubbard model represent a homogeneous system, and so fail to
capture fully the effects of doping.  This suggests that a proper description 
of such doped materials should include the combined effects of correlations 
and disorder on an equal footing.

Actually, the one-band Hubbard model with static, diagonal disorder has 
already been the subject of a few studies.~\cite{a2,a3} Jani$\breve{s}$ 
{\it et al.}~\cite{a2} have studied the effects of site-diagonal disorder 
on the stability of the half-filled Mott-Hubbard antiferromagnet in detail, 
using the QMC algorithm to ``solve'' the $d=\infty$ Hubbard model.  
Miranda {\it et al.}~\cite{a3} have considered localization effects in the 
conduction band DOS and their effects on the distribution of the Kondo    
temperature, $P(T_{K})$, and the resulting non-FL behavior in Kondo alloys.
Dobrosavljevic {\it et al.}~\cite{a3} have considered the role of 
{\it off-diagonal} disorder in local moment formation in disordered metals, 
while Sarma {\it et al.}~\cite{a4} investigated the effects of
disorder in the $d=\infty$ Hubbard model in a simple way to study the s.p 
spectral function; they have also applied their approach to compute the 
angle-integrated photoemission spectra in $SrTiO_{3-\delta}$, and 
have demonstrated that the experimental features are reproduced quite well.  
However, their approach involves treating the effects of disorder by a 
simple weighting procedure~\cite{a4} that is reminiscent of the earlier 
known virtual crystal approximation for alloys~\cite{a5}. More recently, 
Mutou~\cite{a4} has used the same treatment for incorporating disorder in 
the Hubbard model in $d=\infty$. It is known that the VCA is valid for 
small impurity concentration, and does not interpolate correctly between 
the weak- and the strong scattering limits, where one has to deal with 
resonant scattering.  In the latter case, a $T$-matrix approach is more 
suitable. Indeed, it is known from studies of {\it non-interacting},
disordered systems that the so-called coherent-potential approximation (CPA)
solves the Anderson disorder problem exactly in $d=\infty$.~\cite{a6}  The 
effect of disorder in the paramagnetic, strongly correlated FL metallic 
phase of the $d=\infty$ Hubbard model within a framework valid for all 
parameter regimes (disorder strength, interaction strength and filling) 
has not yet been considered.  

  In this paper, we consider the effects of strong correlations (via the local
Hubbard $U$) and static, diagonal disorder in the {\it metallic}, paramagnetic 
state of the $d=\infty$ Hubbard model.  Consequently, disorder induced 
Anderson localization cannot be accessed within our technique, and we will be
interested in the effects of disorder-induced strong scattering on the 
correlated metallic phase away from half-filling.
 
A naive analysis might lead one to conclude that, at least at low energies, 
and in $d=\infty$, disorder would have qualitatively the same effects in a 
free ($U=0$) system and in an {\it interacting} Fermi system on a lattice.  
One might argue that local correlations act to renormalize the bare energy 
scales, keeping the Fermi liquid correspondence intact, and disorder would 
then have qualitatively the same effects as in the free ($U=0$) case.  This 
reasoning, however is not obvious for reasons pointed out by 
Jani$\breve{s}$ {\it et al}.~\cite{a2}  With 
strong, local correlations, a M-I transition is possible in a pure Hubbard 
model. It is known that this transition is characterized by a transfer of 
dynamical spectral weight over large energy scales, a feature that is 
completely absent in non-interacting models.  In the $d=\infty$ Hubbard 
model, for e.g, spectral weight is transferred over large energy scales 
as the temperature is lowered through the lattice Kondo scale, 
$T_{K}^{latt}$,~\cite{a6} giving rise to nontrivial effects in thermodynamics, 
and in dc and ac transport.  This makes emergence of qualitatively new 
behaviors (which have no analogues in the free system) possible, and hence a 
consistent framework to study correlations and disorder on an equal footing 
is in order.  

In this work, we aim to study the combined effects of correlations and 
static, site-diagonal disorder in the quantum paramagnetic phase of the 
$d=\infty$ Hubbard model.  The questions we want to answer are:

(1) Is there an instability of the strongly correlated Fermi liquid metal to 
an incoherent, non-Fermi liquid metal as the disorder is increased, and, if 
so, what strength of disorder is required?

(2) What is the character of this non-Fermi liquid state, if it exists?

(3) What is the critical disorder strength required to drive this metallic
state insulating? (notice that this is more appropriately termed, 
``Anderson-Mott insulator'' following~\cite{a2}). 

(4) What is the effect of the above scenario (if it exists) on the dynamical 
responses, e.g, on the optical conductivity, electronic Raman lineshape, etc?

In this work, we aim to provide answers to these questions within the 
dynamical mean field theory (DMFT, $d=\infty$) of strongly correlated 
fermionic systems.  In the next section, we set up the formalism to study 
the combined effects of correlations and disorder exactly in $d=\infty$.  
This is followed by a detailed discussion of the role of site-diagonal 
disorder in a strongly correlated FL metal, and on the nature of the 
incoherent metallic state at intermediate disorder strength.  In particular, 
the qualitatively different results we obtain (in comparison to~\cite{a4}) 
are emphasized, and the differences are shown to arise from the fact that 
the CPA, in contrast to VCA, treats the weak- and strong scattering limits 
equally well.~\cite{a5} Lastly, as an application, we describe how the 
formalism presented here can be fruitfully applied to study the interplay 
of correlations (competition between atomic and itinerant tendency) and 
static, diagonal disorder on the ac conductivity, and the electronic Raman 
lineshape of doped transition metal compounds.  The manifestation of the
combined effects of the local Hubbard interaction and local, static disorder
on the dynamical charge response of doped, strongly correlated metals has 
not been considered to date, and ours are the first attempt in this direction.

\section{MODEL AND LOCAL SPECTRAL DENSITY IN $d=\infty$}

We start with the one-band Hubbard model (HM) with static, site-diagonal 
disorder on a hypercubic lattice in $d=\infty$,

\be
H = - t\sum_{<i,j>,\sigma}(c_{i\sigma}^{\dag}c_{j\sigma}+h.c) +
U\sum_{i}n_{i\uparrow}n_{i\downarrow} + \sum_{i\sigma}v_{i}n_{i\sigma}
\ee
where the disorder potentials $v_{i}$ are specified by a given probability
distribution.  We restrict ourselves to {\it microscopic} disorder, and so 
work with a binary-alloy distribution for disorder; explicitly,
\be
P(v_{i})=(1-x)\delta(v_{i}) + x\delta(v_{i}-v)\;.
\ee
  In other words, upon doping, a fraction $x$ of the sites have an additional 
local potential $v$ for an electron (or hole) hopping onto that site.  Given 
this, the carriers experience different local environments in the course of    
their hopping, and the physically sensible object is the disorder-averaged
local Green function (GF), \mbox{$<G_{ii}(\omega)>_{c}$}, where $<..>_{c}$ 
means a disorder average. In $d=\infty$, all interesting dynamical 
information is entirely contained in $<G_{ii}(\omega)>_{c}$, the 
computation of which is the central aim of this section. 

As stated above, the problem of interacting, disordered fermions requires a 
simultaneous consideration of both on an equal footing.  To see how 
one proceeds, notice that the model, eqn.(1) is exactly soluble in 
$d=\infty$ by CPA when $U=0$, and approximately by the IPT~\cite{a6} 
when $v=0$.  This is a reliable technique to solve the $d=\infty$ problem, 
and yields results in good agreement with numerical studies in the 
paramagnetic phase. 
To treat disorder and interactions simultaneously requires a 
suitable extension, either from the $U=0$ limit, where the disorder problem 
is first solved exactly, and the interactions are then considered, using the 
CPA DOS as an input to the interacting model, or the disorder-free Hubbard 
model is first solved by IPT in $d=\infty$, and the IPT GF serves 
as an input to the CPA procedure.  In what follows, we take the latter 
route.  In principle, both starting points should yield the same final 
result in a fully selfconsistent treatment of interactions and disorder on 
the same footing.  To be consistent, consideration of the competition between  
interactions in the metallic phase and static disorder requires a fully 
selfconsistent procedure, which we describe in what follows.

We start by writing down the local effective action for the impurity model
with site-diagonal disorder;~\cite{a2}
\beq
\nonumber 
S_{eff}[i] & = & -\sum_{\sigma}\int_{0}^{\beta}d\tau\int_{0}^{\beta}d\tau' 
c_{i\sigma}^{\dag}(\tau)g_{\sigma}^{-1}(\tau,\tau')c_{i\sigma}(\tau') 
\\ & + &  
U\int_{0}^{\beta}n_{i\uparrow}(\tau)n_{i\downarrow}(\tau)
\eeq
where the Weiss fn., describing the dynamics on the rest of the lattice is 
given by,
\be
g_{\sigma}^{-1}(\tau,\tau') = \delta(\tau-\tau')(-v_{i}-\partial_{\tau}) -
t^{2}\int dv_{i}P(v_{i})G_{ii\sigma}(v_{i},\tau,\tau')
\ee
with the integral over $v_{i}$ representing the disorder average.  In the 
above, $G_{ii}$ is the fully interacting local Green fn. of the lattice 
model. To solve the effective problem, one has to first solve the interacting 
problem, followed by a disorder average, and link these up in a proper 
selfconsistent way. In this work, the interaction part is treated via 
selfconsistent iterated perturbation theory (IPT) away from 
half-filling.~\cite{KK} We used this technique as it yields the correct 
Fermi liquid metallic state, and gives results in very good agreement with 
numerical studies in the quantum paramagnetic regime, which we consider 
here.  The interplay between local correlations and static disorder is 
studied by a combination of the IPT with the CPA , as described below.

{\bf Step 1}  The disorder-free Hubbard model, eqn.(1), is ``solved'' in 
$d=\infty$ by iterated perturbation theory at and off half-filling.~\cite{KK}  
This maps the lattice problem onto an effective one-body problem describing 
the propagation of renormalized fermion quasiparticles in a complex effective 
medium, as described by a local self-energy, $\Sigma(\omega)$.  This 
procedure is known to yield the correct FL behavior in the quantum 
paramagnetic phase of the $d=\infty$ Hubbard model, and gives results in 
good agreement from those obtained from exact diagonalization studies.  
The effective one-body Hamiltonian now is formally written as,
\be
H_{eff}=\sum_{\bf k}[\epsilon_{\bf k}+\Sigma(\omega)]
c_{{\bf k}\sigma}^{\dag}c_{{\bf k}\sigma}
\ee
To treat the effects of disorder, we follow the philosophy of the CPA, which 
entails removing one site from the effective medium, and replacing it with the
actual local potential, $v_{i}$.  The {\it effective} local potential that 
scatters the electrons (holes) is now $V_{i}=v_{i}-\Sigma(\omega)$, with
$\Sigma(\omega)$ to be determined selfconsistently from the CPA condition.
This condition requires that the additional scattering produced by the 
effective potential vanish on the average; this is equivalent to the 
requirement that the configuration-averaged $T$-matrix vanish, 
$<T_{ii}[\Sigma(\omega)]>_{c}=0$. In $d=\infty$, the usual procedure is 
to map the lattice problem onto this single-site problem supplemented 
with a selfconsistency condition linking this site to the rest of the 
lattice.  The single-site hamiltonian is,
\beq
\nonumber 
H & = & \sum_{\bf k}[\epsilon_{\bf k}+\Sigma(\omega)]
c_{{\bf k}\sigma}^{\dag}c_{{\bf k}\sigma} + 
\sum_{i,{\bf k},\sigma}t_{\bf k}(c_{{\bf k}\sigma}^{\dag}c_{i\sigma}+h.c) 
\\ & + & 
\sum_{i\sigma}(v_{i}-\Sigma(\omega))n_{i\sigma}
\eeq
for the problem with disorder.~\cite{a6}

{\bf Step 2}  The local spectral density, computed for the pure HM in 
$d=\infty$, is used as an input to the CPA procedure, which computes a 
new self energy which takes account of disorder induced scattering processes 
exactly in $d=\infty$.~\cite{a5,a6} Given the IPT DOS, $\rho(\omega)$, 
from \mbox{Step 1}, the CPA yields the following eqns for the corrected 
self-energy and the disorder averaged GF:

\be
\Sigma_{c}(\omega)=xv+\frac{v^{2}x(1-x)}{\omega-v(1-x)-
\cal{A}(\omega)}
\ee
and,
\be
<G_{ii}(\omega)>_{c}=\frac{1-x}{\omega-\cal{A}(\omega)}
+\frac{x}{\omega-v-\cal{A}(\omega)}
\ee
where $x$ is the concentration of dopants, created by alloying or chemical 
substitution, and $\cal{A}(\omega)$ is the effective dynamical (Weiss) field 
for the model with disorder and/or Hubburd on-site correlations. A complete 
derivation of the above formulae is given in the appendix for convenience.  

{\bf Step 3}  The full, disorder averaged GF, $<G_{ii}(\omega)>_{c}$, is used
as a new input to the IPT routine that recomputes the local self-energy of the 
interacting problem with the modified GFs from the zero-order iteration that
involves steps (1+2).  The modified IPT propagator is then fed back into the 
CPA routine.  The updated GF is then fed back into the IPT routine and the
process is iterated to convergence.

{\bf Step 4}  At each stage, the Luttinger sum rule, 
$n_\sigma=\int_{-\infty}^{\mu}\rho_\sigma (\omega) d\omega$, 
is satisfied to a high accuracy.  The iteration procedure described above 
describes the {\it full} fermion propagator dressed by interactions (IPT) 
and disorder scattering (CPA) and the selfconsistency procedure ensures 
that the mutual interplay of correlations and disorder is treated 
consistently in $d=\infty$.

All dynamical quantities of interest in the quantum paramagnetic metallic 
regime are obtained from $G_{ii}(\omega)$.  The electrical transport 
properties for the DHM can be computed from the spectral density.  
The optical conductivity for the DHM can be computed from the local GFs,
since vertex corrections in the Bethe-Salpeter eqn for the conductivity vanish
identically in $d=\infty$.~\cite{a6}  The relevant eqn is,
\be
\sigma_{xx}(i\omega)=\frac{1}{i\omega}\int \rho_{0}(\epsilon)
\sum_{i\nu} G(\epsilon,i\nu) G(\epsilon, i\omega+i\nu)
\ee

Moreover, in the same approximation, the Raman intensity lineshape is simply
related to the optical conductivity as,~\cite{a7}
\be
I_{xx}(\omega)=\frac{\omega}{1-e^{-\beta\omega}}Re \sigma_{xx}(\omega)
\ee

The complete magneto-optical response of the DHM in the quantum 
para-metallic state can also be readily computed in a way identical 
to that used for the pure HM.~\cite{a8}

\section{RESULTS IN $d=\infty$}

We now proceed to discuss our results.  In our calculations, we employed
a gaussian unperturbed DOS,~\cite{a6} and $U/D =3.0$ ($D$ is the free 
bandwidth). We used the IPT off $n=1$ as a reliable approximation to treat 
the correlation part exactly in $d=\infty$.  The IPT gives results in 
excellent agreement with exact diagonalization studies,~\cite{a6} and 
we have checked that all desirable FL properties are reproduced in our 
numerics.  We restricted ourselves to the paramagnetic metallic phase 
throughout. We varied the disorder strength in the range $0 \le v \le U$ 
at fixed $n$, and repeated our calculations for each 
$n=1,~0.9,~0.8$ to get a clear picture of the combined effects of 
(correlations + disorder) and band-filling.
\begin{figure}[t]
\epsfxsize=3.5in 
\epsffile{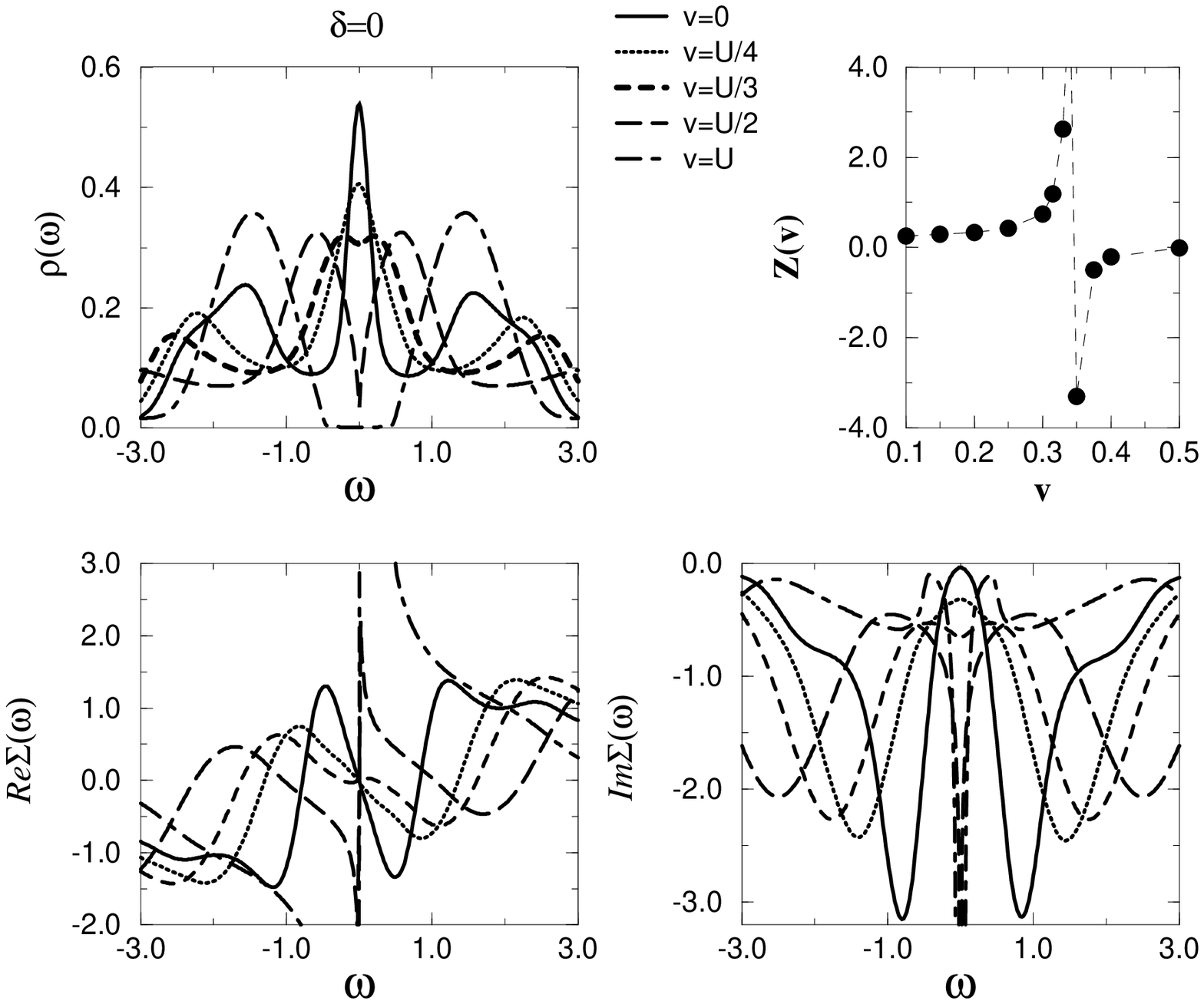} 
\caption{Local spectral density, and real and imaginary parts of the s.p 
self-energy for the disordered Hubbard model in $d=\infty$.}
\label{fig1}
\end{figure}

Fig.~\ref{fig1} shows the result of the calculation for the local DOS of 
the DHM in $d=\infty$.  For $v=0$, the DOS shows all the characteristic 
features~\cite{a6} of the strongly correlated FL metal (also shown up in the 
self-energy, the real and imaginary parts of which are also shown in 
fig.~\ref{fig1}).

\begin{figure}[b]
\epsfxsize=3.5in 
\epsffile{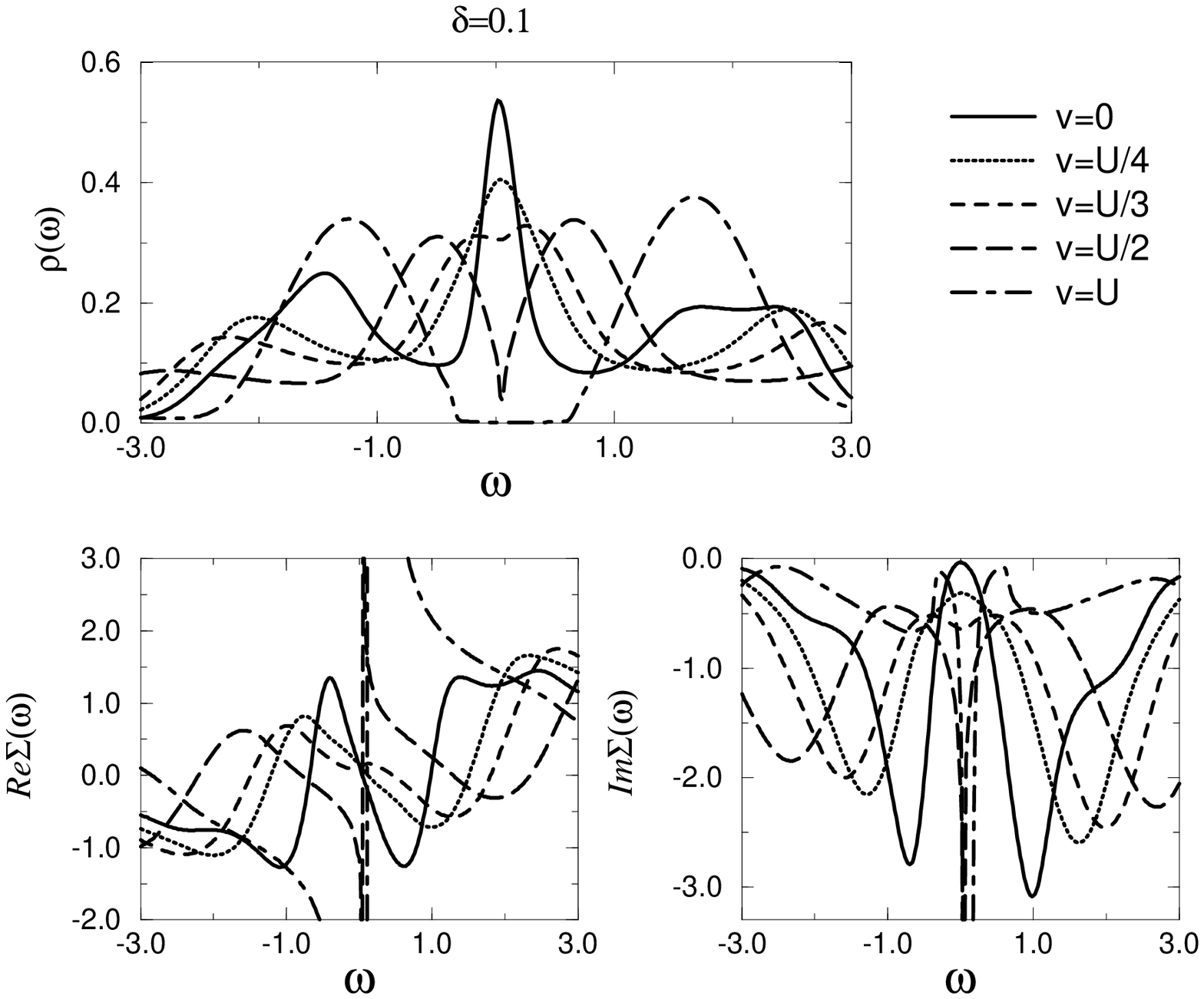} 
\caption{Local spectral density, and real and imaginary parts of the s.p 
self-energy for the disordered Hubbard model in $d=\infty$.}
\label{fig2}
\end{figure}

The dotted line in fig.~\ref{fig1} shows the local spectral density 
and the self-energy for small ($v=U/4$) disorder strength.  As expected, the 
central FL peak is broadened by disorder, and 
Im$\Sigma (\omega)=-a-b\omega^{2}$ acquires a finite value at the Fermi 
surface due to disorder scattering. Notice that the height of the central 
peak has decreased from its non-interacting value for $v=U/4$, and so 
Luttinger's theorem is not satisfied (notice that the Luttinger sum rule, 
fixing the value of the chemical potential, is still satisfied, but 
Luttinger's theorem breaks down, in the sense that a sharp Fermi surface 
no longer exists).  This is the consequence of the ${\bf k}$-space degeneracy 
arising from the introduction of disorder.  Nevertheless, the self-energy 
shows a behavior characteristic of a {\it dirty} Fermi liquid.  We notice 
that an analogous situation would arise in a treatment of the disorder in 
the Born approximation (or its selfconsistent versions), and refer to this 
metallic state as a disordered Fermi liquid.  At $v=U/3$, however, one sees 
the emergence of a qualitatively new behavior in fig.~\ref{fig1} (dashed 
line); a pseudogap develops in a {\it continuous} way 
in the local DOS, and $\Sigma (\omega)$ has a {\it qualitatively} 
new behavior.  Im$\Sigma (\omega)$ has a {\it minimum} at $\mu$, and 
Re$\Sigma (\omega)$ has a positive slope, invalidating the quasiparticle
picture.  We refer to this phase as the {\it incoherent} metallic phase.  
With increasing disorder strength, the pseudogap in the DOS deepens, and 
the shallow minimum in Im$\Sigma (\omega)$ develops into a sharp peak 
at $\mu$ (fig.~\ref{fig1} with the long dashed line).  Finally, at $v=U$, 
a clear gap develops at $\mu$, and the incoherent metallic state undergoes 
a {\it continuous} transition to an ``Anderson-Mott'' insulator. Similar 
trends are observed off $n=1$, as shown in fig.~\ref{fig2} for $n=0.9$.  

The quasiparticle residue, $Z(\mu)$, calculated as $Z(\mu)=1/(1-F(\mu))$, with
$F(\mu)=d[Re\Sigma(\omega)]/d\omega|_{\omega=\mu}$, illustrates the FL- non FL
"transition" in a clearer way (fig.~\ref{fig1}).  With $U/D=3.0$, $n=1$, the 
Fermi liquid is stable upto $v/U=(v/U)_{c} \simeq 0.31$.  For $v > v_{c}$, we 
see that $Z(\mu) > 1$, or even negative, invalidating the very concept of 
coherent FL quasiparticles.  A similar behavior of $Z(\mu)$ is obtained within 
the CPA treatment of the non-interacting, disordered model (see appendix). 
This illustrates how the FL metal is destroyed {\it continuously} with 
increasing disorder before the Anderson-Mott insulating phase is approached 
around $v/U \simeq 1$.  This insulating phase results from the complete 
splitting of the DOS for large $v$, and happens in a way similar to that 
in usual CPA. It is not caused by Anderson localization, which cannot be 
accessed in $d=\infty$, as is known.  Thus, for small values of $v$, 
the ``FL metal'' 
survives; for intermediate $v$, we have shown that the metallic state is 
{\it not} a Fermi liquid; it is rather an incoherent metal. This incoherent 
metal is unstable to a Mott-Anderson insulator beyond a certain $v=v_{c}$.  
These results are very different from those obtained in the treatment of 
the non-interacting, disordered model using CPA. Indeed, in this case, the 
incoherent metal phase develops for {\it any} value of the disorder. To see 
this, we use the CPA eqns for the selfenergy with $U=0$, whereby 
$G(\omega)=G_c(\omega)$. The CPA yields,
\be
\Sigma_{c}(\omega)=xv+\frac{v^{2}x(1-x)}{\omega-v(1-x)+iD^{2}\pi\rho_{0}(\mu)}
\ee
as shown in the appendix.  With a positive definite $\rho_{0}$, 
Re$\Sigma_{c}(\omega)$ has a positive slope near $\mu$ (see appendix), 
and so the metallic state is not a FL.  The above illustrates the importance 
of treating both correlations and disorder on an equal footing, and shows 
that the actual behavior in such systems is quite different from those with 
only interactions (no disorder), or those with only disorder (no 
interactions). 

The above results are also quite different from those obtained from a VCA 
treatment of disorder.~\cite{a4} To see this, we have repeated the calculation 
carried out by Mutou, and have observed that the incoherent, non-FL 
pseudogap metal phase never sets in, no matter how strong the disorder.  
This is a direct consequence of the fact that the VCA cannot treat resonant 
scattering correctly (intermediate $v$) while it is known that the CPA 
works equally well for {\it all} values of disorder.~\cite{a2}  In fact, our 
results (fig.~\ref{fig1}, fig.~\ref{fig2}) are completely consistent 
with those of Mutou~\cite{a4} when $v$ is small ($v=U/4$), in which case, 
the CPA as well as the VCA give results in accordance with a 
perturbative treatment of disorder.

\section{TRANSPORT PROPERTIES OF THE DHM IN $d=\infty$}

\begin{figure}[htb]
\epsfxsize=3.5in 
\epsffile{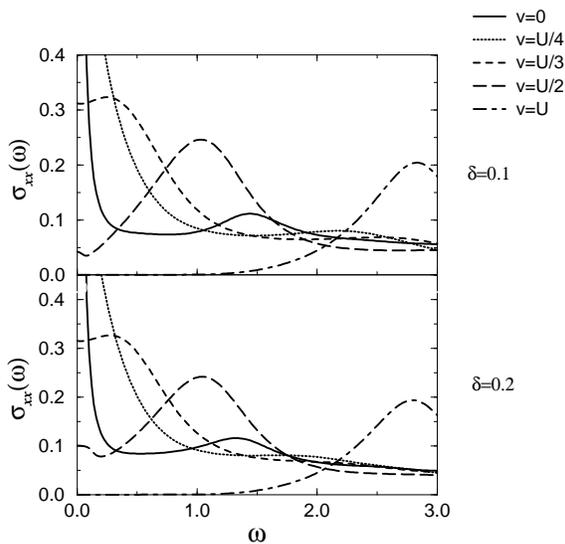} 
\caption{Optical conductivity of the disordered Hubbard model in $d=\infty$
as a fn. of disorder strength for two different band-fillings, $n=0.9, 0.8$.}
\label{fig3}
\end{figure}

  Knowledge of the full disorder Green function is a sufficient input
for a controlled computation of transport properties in $d=\infty$.  This is
because vertex corrections drop out in the Bethe-Salpeter eqn. for the 
two-particle propagator, for e.g, for the conductivity, and the 
corresponding susceptibility is just the convolution of the {\it full} 
GFs.~\cite{a6}  Using the formulae given in Section II, we have 
computed the optical conductivity and the electronic Raman scattering 
lineshape for the DHM.

\begin{figure}[htb]
\epsfxsize=3.5in 
\epsffile{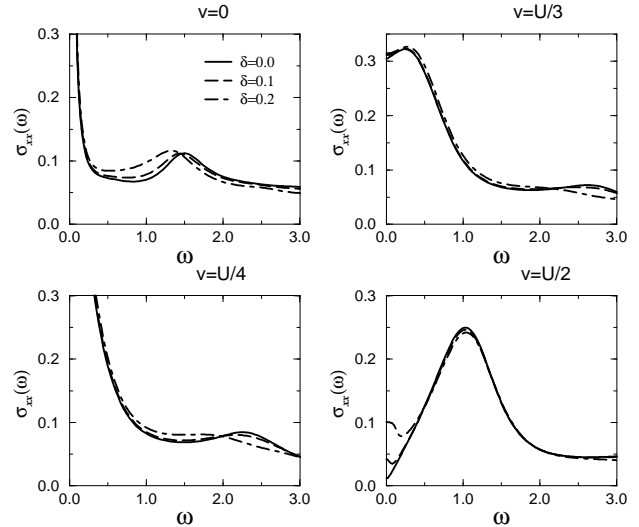} 
\caption{Optical conductivity of the disordered Hubbard model in $d=\infty$
as a fn. of band-filling for different disorder strengths: 
$v/U=0, 1/4, 1/3, 1/2$.}
\label{fig4}
\end{figure}

The changes in the local spectral density as a function of $v$ and filling
for a given $U/D$ are reflected in the optical spectra. fig.~\ref{fig3} 
shows the calculated optical conductivity for $U/D=3.0$, $n=0.9,~0.8$ and for
different disorder strengths.  The solid line corresponds to the pure Hubbard
model; we see that all expected features are reproduced in good agreement 
with earlier work.~\cite{a6}  In particular, the ``Drude'' peak at 
$\omega=\mu$, the mid-IR peak around $U/2$, which corresponds to transitions 
between the lower-Hubbard band and the central FL resonance, and the 
high-energy peak corresponding to transitions between the Hubbard bands, are 
all reproduced well.  As expected from the evolution of the DOS, small 
disorder, $v=U/4$, broadens the ``Drude'' peak, and smoothens out the 
mid-IR and high energy features, while preserving a FL response (dotted 
line).  With $v=U/3$, however, all semblance of the Drude peak has 
disappeared completely, and the pseudogap formation in the DOS is clearly 
reflected in $\sigma(\omega)$.  The dashed line ($v=U/3$) therefore 
represents the optical response characteristic of an incoherent non-FL metal
(since the DOS at $\mu$ is finite).  Notice that the mid-IR feature is 
progressively broadened and shifted to higher energies with increasing $v$ 
as seen for $v=U/2$ (long dashed line). Finally, the onset of the 
Mott-Anderson insulating phase is clearly reflected in the appearance of 
a threshold feature in the optical response.  This scenario is only slightly 
modified as a function of band-filling, as shown in fig.~\ref{fig3}.  

It is known that the optical response of the Hubbard model~\cite{a7} shows up 
the dramatic transfer of spectral weight from high- to low energy on hole 
doping, a characteristic of strongly correlated systems.  A very interesting, 
related observation is that the $\sigma_{xx}(\omega)$ curves for different 
hole doping cross at a single point, $\omega_{c}$, the so-called 
{\it isosbectic} point.  Similar features are observed for the DHM as a fn. 
of hole-doping, as seen in fig.~\ref{fig4}.  We observe that the 
$\sigma_{xx}(\omega)$ curves now seem to cross at {\it two} points
as a fn. of doping.  This seems to be true for all disorder strengths, and
clearly warrants a closer examination.  Heuristically, we can understand the
crossing of the curves as arising from the high- and low-frequency behavior of
the charge susceptibility;~\cite{a7} a microscopic understanding of the two 
crossing points is, however, a harder task.  

The qualitatively different behavior of the
optical response for varying disorder strengths can be understood in terms of 
the coherent (incoherent) part of the low-frequency spectral density.  At zero
or weak disorder, the Green fn. still is that characteristic of a Fermi liquid
(real or complex pole structure), and $\sigma_{xx}(\omega)$ shows a ``Drude''
peak at $\omega=\mu$.  With increasing disorder, one is in the resonant 
scattering regime, all semblance of the quasiparticle behavior is destroyed, 
and the GF is characterized by a branch-cut behavior.  The difference in the 
spectra as a fn. of disorder is then a consequence of the fact that the 
response, which is caused by the action of the current operator (which 
has non-zero matrix elements between lower-Hubbard band states) does not 
create well-defined elementary excitations as $v$ is increased.  Within 
the $d=\infty$ ideas used here, the coherent response is the manifestation 
of the collective band Kondo effect for zero or weak disorder, while the 
incoherent response for larger $v$ is a consequence of the suppression of 
this coherence scale by disorder-induced strong scattering.

\begin{figure}[htb]
\epsfxsize=3.5in 
\epsffile{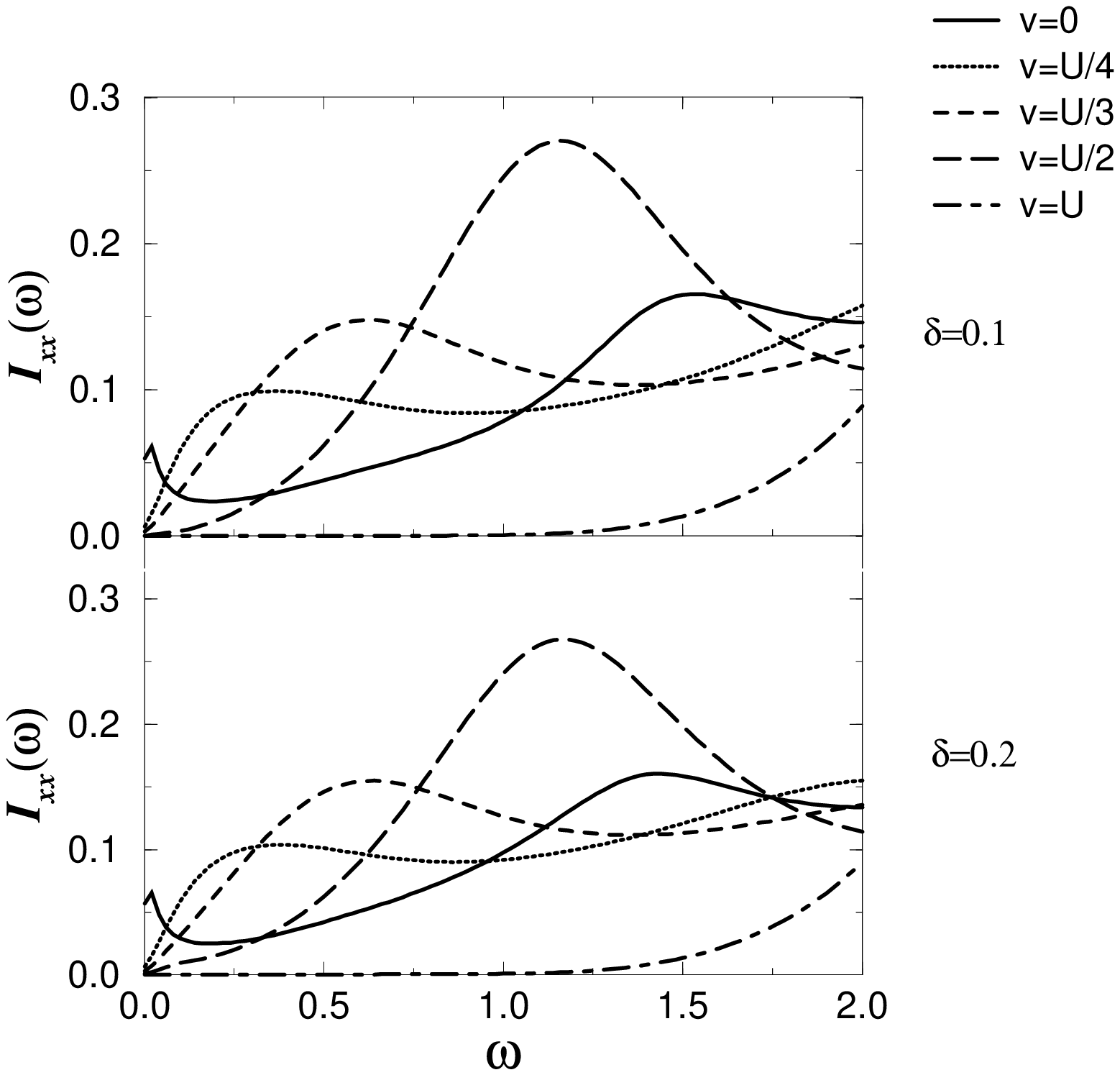} 
\caption{Electronic Raman scattering lineshape for the disordered Hubbard 
model in $d=\infty$ as a fn. of disorder for two values of band-filling,
$n=0.9, 0.8$.}
\label{fig5}
\end{figure}

In $d=\infty$, the electronic Raman scattering lineshape can be directly
obtained from the optical conductivity, as remarked above.  In light of the 
above discussion, we expect a sharp electron-hole peak, characteristic of 
coherent particle-hole response in a FL, for $v=0$, and a disorder induced 
broadening of this feature for small $v=U/4$.  For larger $v$, we expect a 
completely incoherent lineshape characteristic of the non-FL pseudogap 
metal state ($v=U/3,~U/2$).  This is indeed what we observe in our 
calculations (fig.~\ref{fig5}) and is related to the fact that the 
collective particle-hole model is overdamped by disorder-induced 
resonant scattering in this regime.  The evolution of the 
Raman spectrum with increasing disorder is then understood as follows.  The
stress tensor (whose fluctuations are measured in the Raman lineshape), which
connects the eigenstates of the Hubbard Hamiltonian, creates well defined 
elementary excitations at low energy in a FL, giving a sharp peak at low 
energy, while in the incoherent metal regime, the Green function has a 
branch cut, and the action of the stress tensor does not create 
well-defined elementary excitations, resulting in a continuum response 
at low energy.

\begin{figure}[htb]
\epsfxsize=3.5in 
\epsffile{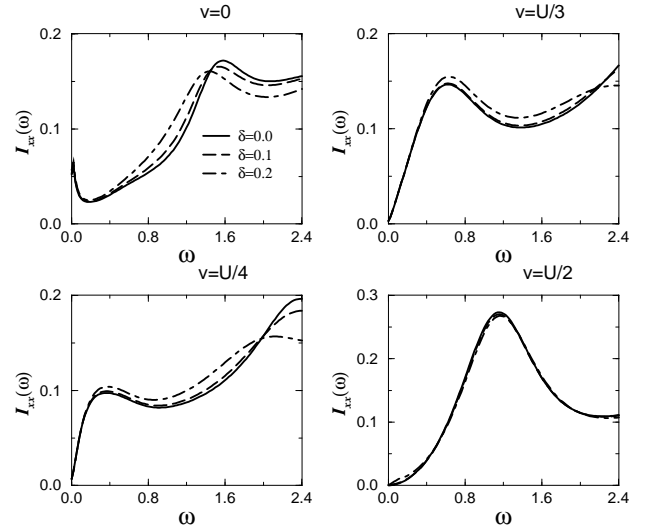} 
\caption{Electronic Raman scattering lineshape for the disordered Hubbard 
model in $d=\infty$ as a fn. of band-filling for different disorder strengths, 
$v/U=0, 1/4, 1/3, 1/2$.}
\label{fig6}
\end{figure}

The transfer of spectral weight with doping is also revealed in the Raman 
lineshapes (fig.~\ref{fig6}), and is understood directly in terms of the
evolution of $\sigma_{xx}(\omega)$ as a fn. of doping.  In fig.~\ref{fig6}, 
we show the Raman lineshapes as a fn. of doping for four different
disorder strengths, $v=0,~U/4,~U/3$ and $v=U/2$.  In spite of the 
different nature of the metallic state for the first two disorder values, 
the isosbectic point in $I_{xx}(\omega)$ is clearly revealed, seemingly 
independent of details of the low-energy spectra.  

In conclusion, we have considered the role of static, site-diagonal disorder
in the quantum paramagnetic phase of the $d=\infty$ Hubbard model.  We have
devised a formalism that captures the interplay of dynamical local correlations
inherent in the Hubbard model and doping-induced disorder on an equal footing.
This is achieved by a proper extension of the IPT off half-filling, which has
been shown to work well in the paramagnetic phase of the $d=\infty$ Hubbard
model, by combining it (selfconsistently) with the CPA, which is known to 
yield the exact solution of the Anderson disorder problem in $d=\infty$.  We
have shown that the FL metal is stable against {\it small} disorder ($v=U/4$),
and that strong repeated scattering effects at higher disorder values destroys
the FL coherence, giving rise to an incoherent pseudogap metal phase.  At a 
critical $v$, this phase becomes unstable to a Mott-Anderson insulating phase.
We have also considered the effects of the interplay between correlations in 
the FL metallic state and disorder in $d=\infty$ on the optical and Raman 
response, and have shown how low-energy coherence (incoherence) is manifested 
in the dynamical charge response of correlated, disordered systems in this
limit.  Lastly, consideration of effects akin to Anderson localization require
an extension of the approach to include non-local correlations, or extensions 
along the lines of Miranda {\it et al}.~\cite{a3}  We plan to study such 
effects in the future.

\section{APPENDIX}

In this appendix, we review briefly the CPA in the interaction-free case 
($U=0$).  The CPA essentially involves (i) replacing the actual disorder 
problem by an {\it effective} one-body problem, where the electrons now 
move in a dynamical (complex) effective medium.  (ii) One now removes this 
complex potential (self energy, $\Sigma(\omega)$) from one local site, $i$, 
and replaces it with the actual local potential, $v_{i}$.  The additional 
effective potential seen by an electron is now $V_{i}=v_{i}-\Sigma(\omega)$.  
For a binary alloy disorder, $P(v_{i})=(1-x)\delta(v_{i})+x\delta(v_{i}-v)$, 
one needs to average additionally over disorder.  The site-diagonal 
$T$-matrix corresponding to this additional effective scattering potential 
$V_{i}$, after configuration averaging, is set to zero in the CPA 
condition:~\cite{a5} 
\be
\frac{(1-x)(-\Sigma(\omega))}{1+\Sigma(\omega)G(\omega)} + 
\frac{x(v-\Sigma(\omega))}{1-(v-\Sigma(\omega))G(\omega)}=0 
\ee
Solving for the self-energy, $\Sigma(\omega)$, using 
$G^{-1}(\omega)=\omega-\Sigma(\omega)-\cal{A}(\omega)$ gives the 
eqns.~(7) and~(8) used in the text.

Next, we show that the unusual feature of the $Z(\mu)$ vs $v$ (fig.~(1)) can 
be understood analytically in the CPA as well.  In fact, as we show below, 
the CPA in the case with $U=0$ always leads to $Z(\mu)>1$, showing that the 
metallic state is never a Fermi liquid.  Choosing the unperturbed DOS to be 
a lorentzian with half-width $D$, one can show~\cite{a6} that 
${\cal A}(\omega)=-iD$. The selfconsistent eqns. now reduce to a single 
algebraic eqn. for the self-energy, giving exactly eqn. (11) in the text 
with $\rho_{0}(\mu)=1/D\pi$.  The resulting real part of the self-energy has 
a {\it positive} slope at low energy, giving $Z(\mu)=[1-(v/2D)^{2}]^{-1}$, 
which is never FL-like.

\acknowledgments
One of us (MSL) thanks SFB341 for financial support.  
LC acknowledges financial support of Funda\c c\~ao de Amparo \`a 
Pesquisa do Estado de S\~ao Paulo (FAPESP).

\end{document}